\documentstyle[aps,amsmath]{revtex}

\begin{document}

\title{Langevin equation for the extended Rayleigh model with an asymmetric bath}
\author{Alexander V. Plyukhin and Jeremy Schofield\\
{\small {\sl Chemical Physics Theory Group, Department of Chemistry,}}\\
{\small {\sl University of Toronto, Toronto, Ontario, Canada M5S 3H6}}}
\date{\today}
\maketitle

\begin{abstract}
In this paper a one-dimensional model of two infinite gases 
separated by a movable heavy piston is considered. The non-linear 
Langevin equation 
for the motion of the piston is derived from first principles 
for the case when the thermodynamic parameters and/or the molecular 
masses of gas particles on left and right sides of the piston are
different. Microscopic expressions involving time correlation
functions of the force between bath particles and the piston are
obtained for all parameters appearing in 
the non-linear Langevin equation.  It is demonstrated that
the equation has stationary solutions
corresponding to directional
fluctuation-induced drift in the absence of systematic forces.
In the case of ideal gases interacting with the piston via a  
quadratic repulsive potential, the model is exactly solvable and
explicit expressions for the kinetic coefficients in the non-linear
Langevin equation are derived.  The transient
solution of the non-linear Langevin equation is analyzed
perturbatively and it is demonstrated that previously obtained results
for systems with the hard-wall interaction are recovered.

\end{abstract}

%\begin{multicols}{2}

\section{Introduction}

The Brownian motion of a massive piston in a cylinder filled with 
an ideal gas   
is one of the oldest models of non-equilibrium statistical physics.
In its simplest version, both the piston (of mass $M$) and the gas molecules
(of mass $m$) are confined to move in 
one dimension along the symmetry axis of the cylinder.
The piston is assumed to be adiabatic in a sense that
its heat conductivity is negligible so that there is no transfer of heat
between the two compartments of the cylinder as long as the piston is held
fixed.  This simple model has served as a useful test of
new schemes and approximations for several generations of physicists
(see Ref.~\cite{PS} for references).  
In most treatments it is assumed that the bath particles interact with 
the piston via the hard-wall potential, i.e. piston-molecule collisions 
are considered to be instantaneous.
In this model the piston interacts with bath
particles through a sequence of binary collisions and the possibility
of simultaneous interaction of the piston with more than one molecule 
is neglected. This approximation is valid as long as one is interested
in the asymptotic long time behavior of the system 
which is apparently insensitive to details of the
interactions between the bath particles and the piston.
On the other hand, the motion of   
the system on short to intermediate time scales is clearly influenced
by the nature of bath particle-piston interactions, and one has to
take into account the finite range of the potential and 
effects of multi-particle collisions.  
It is known that many-particle collisions are an important factor 
in liquid-like systems, essentially affecting the shape of the velocity 
correlation function, especially when a repulsive part 
of potential is relatively ``soft''~\cite{cage}. In such systems, a
given particle interacts simultaneously with a relatively heavy 
swarm of other particles in its vicinity, leading to an enhancement of caging 
effects in comparison to systems with a short-ranged potential.  
As a result, 
the velocity autocorrelation function 
has a more pronounced negative part corresponding to an
anti-correlation in the velocity induced by a particle rebounding of
its neighbors. Multi-particle collisions may also be important 
for tagged particle motion in a dilute gas-like medium  provided  the tagged
particle (or ``piston'' in the present case) 
is large enough that the average number $N$ of gas molecules 
in the interaction shell around the particle is larger than one. 
Although in this case the qualitative form of correlation functions 
does not depend
on $N$, the softness of the potential
essentially influences the short-time behavior of the
system leading to the non-exponential initial decay of time correlation 
functions~\cite{gaussian_decay}. 

Another reason why one may wish 
to go beyond the approximation of hard-wall interaction is 
methodological. It is often problematic to directly apply general   
analytical techniques involving differential operators for systems
with a singular potential. Even if  one expects multiple collisions 
to be unimportant, it is often convenient to model interactions
via a short-ranged potential of range $l_i$ and subsequently analyze
results in the hard-wall limit $l_i \rightarrow 0$.

In a earlier paper~\cite{PS}, a microscopic derivation of the
non-linear Langevin equation was presented for a system consisting of a massive
piston interacting with a bath particles via an arbitrary repulsive potential.
We use the term ``non-linear'' to refer the Langevin 
equation with not only the linear dissipative term
(Stokes damping) but also dissipative terms of higher orders in particle's 
momentum.
In Ref.~\cite{PS}  all generalized kinetic coefficients appearing the
non-linear Langevin equation were expressed
in terms of time correlation functions involving the interaction force
between the piston and the bath particles
and its derivatives. For some systems, such as an ideal gas
interacting with the piston through a repulsive quadratic
potential,  the microscopic expressions for all the kinetic
coefficients can be calculated analytically.  For this reason, simple
models, such as the ideal gas-piston system, provide a convenient means
of studying many subtle points of Brownian motion theory,
such as the role and form of non-linear damping, relative importance of
non-Markovian effects, and convergence properties of small parameter 
expansions.

In Ref.~\cite{PS} the dynamics of the piston in a homogeneous bath was
examined.
The purpose of this paper is to  extend the analysis in Ref.~\cite{PS} 
to the case of an asymmetric bath in which the
parameters characterizing the bath to the left and to the right of the piston 
are different.
This model has received a recent renewal of interest after it was
discovered that the system may exhibit non-trivial 
transient and stationary behavior when there is an initial asymmetry
in the thermodynamics properties of the gases to the left and to the right
of the piston~\cite{general}. 
Among other interesting points,
it was found that the piston undergoes a noise-induced directional 
movement in the absence of macroscopic forces, a characteristic of
motion in molecular motors and stochastic ratchets. 
In the limit of an infinitely long cylinder, Gruber and Piasecki 
found~\cite{GP} a stationary solution of the equations of motion of
the piston corresponding to the drift of the piston
in the direction of the compartment 
with higher temperature 
even when the pressure on the left and the right of the piston are the same.
This result
has been later confirmed by numerical simulations~\cite{experiment}. 
This fluctuation-induced motion
is an effect of the first order in a small mass-ratio 
parameter $\lambda$, defined as $\lambda=\sqrt{m/M}$. It does not follow from
the conventional Langevin equation with linear dissipation, 
and appears only when non-linear
damping terms, which are of higher orders in $\lambda$, 
are taken into account.

In this paper the projection-operator method applied by Mazur and
Oppenheim~\cite{MO} to the theory of Brownian is adapted to the
case of the asymmetric bath. 
We shall consider a slightly more general model of the asymmetric bath
than that analyzed  
by Gruber and Piasecki in that not only temperatures but also 
the masses of the bath particles in the left and 
right compartment may differ. 
For such a system, the Langevin equation,
including non-linear dissipative terms to third order in $\lambda$, is
derived from first principles.  The resulting equation is
general, holds for arbitrary interactions between the piston and
the bath particles, and is not restricted to the ideal gas bath. 
The kinetic coefficients appearing in the non-linear Langevin equation
are expressed as integrals of time
correlation functions of the  force between the piston and bath particles. 
For a bath of ideal gas particles interacting with the piston via 
the parabolic repulsive potential, the correlation functions
can be computed in closed analytical form.
For this model, 
an explicit expression describing relaxation of the momentum of the
massive piston is obtained.  The analysis based on the Langevin
equation is much more simple than that involving 
the language of distribution functions adopted in other
papers on the subject. 
The perturbative and stationary solutions
are analyzed and demonstrated to be consistent with 
the results of Gruber and Piasecki obtained for a model of 
instantaneous binary collisions.

\section{An exact equation of motion for the piston}
The system consists of a piston of mass $M$ and cross-sectional area
$S$ confined to move in one dimension (chosen to be along the $x$
axis) in a  cylinder of total length $L$.  It is assumed that the
piston is initially at position $X$ near the center of the cylinder,
taken to be the origin, with the leftmost and rightmost-ends of the cylinder at
positions $-L/2$ and $L/2$, respectively.  The left and right
compartments of the cylinder are filled with bath molecules of
density $n_l$ and $n_r$, temperature $T_l$ and $T_r$, and mass $m_l$
and $m_r$, and move in one dimension along the $x$ axis.  We assume the
bath molecules are confined to their respective compartments by
hard-wall interactions with the immobile ends of the cylinder, and
interact with the piston through a short-ranged, repulsive  potential.  
The Hamiltonian of the piston-bath system can be written in the form
\begin{eqnarray}
H=\frac{P^2}{2M}+H_0(X),
\label{H1}
\end{eqnarray}
where $X$ and $P$ are the coordinate and the moment of the piston, 
and $H_0(X)$ is the Hamiltonian of the bath in the presence of the  
piston constrained at position $X$.  $H_0(X)$ can be decomposed into 
the sum of two parts, 
$H_0(X)=H_{0}^l(X)+H_{0}^r(X)$,
corresponding to the Hamiltonian for the bath molecules on the left 
and right of the piston,
\begin{eqnarray}
H_0^{\alpha}(X)=\sum_{i=1}^{N_\alpha}\left\{\frac{p_i^2}{2m_\alpha}+
U(X-x_i)\right\}.
\label{H2}
\end{eqnarray}
For simplicity of notation, above and throughout this paper the
super- or subscript index $\alpha=\{l,r\}$ is used to label dynamical
variables in the left and right compartments of the cylinder.
In the above equation $N_\alpha$ and $m_\alpha$ are the number and
the mass of particle in a respective compartment. 

In the following, we restrict our analysis to the case when differences in the 
thermodynamic parameters of the bath in the left and right
compartments of the cylinder are small.  Furthermore, the molecular
masses $m_l$ and $m_r$ are assumed to be of the same order of
magnitude and much less than the mass of the piston $M$. 
Under these conditions, one might anticipate that 
the directional contribution to the momentum of the piston is small, and
on average is of order $P\sim\sqrt{M k_BT_p}$, where the effective 
temperature $T_p$ of the piston 
is of the same order of magnitude as the left and right temperatures
$T_l$ and $T_r$.  In the subsequent analysis, we show that this
intuition is correct, and an explicit expression for $T_p$ will be
presented.

It is convenient to express the equations of motion of the piston-bath
system in scaled coordinates.  To
this end, we introduce the small parameter $\lambda=\sqrt{m/M}$,  where $m$ 
is an arbitrary mass of the same order of magnitude as $m_l$ and $m_r$.  
One  may reasonably  expect that the scaled  momentum of the piston
$P_*=\lambda P$ will typically be of the same order of magnitude
as the momentum of a bath molecule.  The parameter $\lambda$
therefore serves as a quantitative measure of the time scale 
separation between the slow evolution of the massive piston and the
fast evolution of the light bath molecules.  As in the case of Brownian
motion~\cite{MO}, it will be useful as an
expansion parameter to simplify the physics of system~\cite{correction}. 

In terms of the scaled piston's momentum $P_*=\lambda P$, the Hamiltonian
reads $H=P_*^2/2m+H_0$. The corresponding Liouville operator ${\mathcal L}$,
which governs the evolution of an arbitrary dynamical variable via the
equation $A(t)=e^{{\mathcal L}t}A$ ($A\equiv A(0)$ throughout the text), 
can be written in the form
\begin{eqnarray}
{\mathcal L}&=&{\mathcal L}_{0}+\lambda {\mathcal L}_{1},
\,\,\,{\mathcal L}_{0}={\mathcal L}_{0}^l+{\mathcal L}_{0}^r.
\end{eqnarray}
Here the Liouville operators ${\mathcal L}_{0}^{\alpha}$
\begin{eqnarray}
{\mathcal L}_{0}^{\alpha }=
\sum_i \left( \frac{p_i}{m_\alpha}\frac{\partial}{\partial x_i}+
F_i\frac{\partial}{\partial p_i} \right) ,
\end{eqnarray}
describe dynamics of the bath molecules in the left and right
compartments of the cylinder in the presence of the piston held fixed
at position $X$, where the force $F_i$ acting on a molecule $i$
is $F_i=-\partial U(X-x_i)/\partial x_i$. The Liouville operator
${\mathcal L}_{1}$ is defined by
\begin{eqnarray}
{\mathcal L}_1=\frac{P_*}{m}\frac{\partial}{\partial X}+
F\frac{\partial}{\partial P_*},
\end{eqnarray}
where $F=-\sum_i\partial U(X-x_i)/\partial X$ 
is the total force exerted on the piston by the bath molecules. 

Our task is to express the equation of motion of the scaled momentum
of the piston, $dP_*(t)/dt=\lambda F(t)$, in the
form of a Langevin equation in which the instantaneous force $F(t)$
acting on the piston at time $t$ is decomposed into ``dissipative'' and 
``random'' parts.
To accomplish this goal, we define the projection operators ${\mathcal P}_l$, 
${\mathcal P}_r$  
and $\mathcal P$, which act on an arbitrary dynamical variable $A$ according 
to the rules       
\begin{eqnarray}
{\mathcal P}_\alpha A&=&\langle A\rangle_\alpha=
\int d\Omega_\alpha\,\rho_\alpha A,\\
{\mathcal P}A&=&{\mathcal P}_l{\mathcal P}_rA=\langle A\rangle 
=\int d\Omega_l\,\rho_l\int d\Omega_r\,\rho_r\, A ,
\end{eqnarray}
where $\Omega_\alpha=\{x_i^\alpha,p_i^\alpha\}$ are the phase points
for the bath molecules in the left and right compartments of the cylinder,
$\rho_\alpha=Z_\alpha^{-1}exp\{-\beta_\alpha H_0^\alpha(X)\}$
are the corresponding canonical distributions for the bath molecules in
the presence of the piston fixed at coordinate $X$,
$\beta_\alpha = (k_BT_{\alpha})^{-1}$ and $k_B$ is the Boltzmann's constant.  
Note that the
projection operator ${\mathcal P}$ effectively averages over initial conditions of
the bath at a fixed position of the piston.  
Using the operator identity
\begin{equation}
e^{{\mathcal(A+B)}t}=e^{{\mathcal A}t}+\int_0^t d\tau 
e^{{\mathcal A}(t-\tau)}{\mathcal B} e^{({\mathcal A+B})\tau},
\label{operators}
\end{equation}
with ${\mathcal A}={\mathcal L}$ and ${\mathcal B}=-{\mathcal P}{\mathcal L}$, 
one obtains the following decomposition for the force on the piston 
$F(t)=e^{{\mathcal L}t}F$,
\begin{eqnarray}
F(t)=F^\dagger(t) +\int_0^t d\tau\,\,
e^{{\mathcal L}(t-\tau)}{\mathcal P}{\mathcal L}F^\dagger(\tau),
\label{f1}
\end{eqnarray}
where $F^\dagger(t)=e^{{\mathcal Q}{\mathcal L}t}F$ and 
${\mathcal Q}=1-{\mathcal P}$.

The structure
${\mathcal P}{\mathcal L}F^\dagger(\tau)$ in Eq.(\ref{f1})
can be further simplified to $\lambda{\mathcal P}{\mathcal L}_1F^\dagger(\tau)$
due to the orthogonality property,
\begin{eqnarray}
{\mathcal P}_l{\mathcal L}_0^l={\mathcal P}_r{\mathcal L}_0^r=
{\mathcal P}{\mathcal L}_0=0,
\label{property}
\end{eqnarray}
which follows from the relation ${\mathcal L}_0^\alpha \rho_\alpha=0$.
Using this relation and the definition of the projection operators,
Eq. (\ref{f1}) takes the form
\begin{eqnarray}
F(t)=F^\dagger(t) +\lambda\int_0^t d\tau\,\,
e^{{\mathcal L}(t-\tau)}\left\{\frac{P_*}{m}
\left\langle\frac{\partial}{\partial X}F^\dagger(\tau)\right\rangle+
\frac{\partial}{\partial P_*}\langle FF^\dagger(\tau)\rangle
\right\} .
\label{f3}
\end{eqnarray}
In this expression the factor 
\begin{eqnarray}
\left\langle\frac{\partial}{\partial X}F^\dagger(\tau)\right\rangle=
\int d\Omega_l\,\rho_l\int d\Omega_r\,\rho_r\,
\frac{\partial}{\partial X}F^\dagger(\tau) ,
\label{aux1}
\end{eqnarray} 
can be simplified by pulling the differential 
operator out of the integral,
\begin{eqnarray}
\left\langle\frac{\partial}{\partial X}F^\dagger(\tau)\right\rangle=
\frac{\partial}{\partial X}\left\langle F^\dagger(\tau)\right\rangle
-\int d\Omega_l\int d\Omega_r\,F^\dagger(\tau)
\frac{\partial}{\partial X} (\rho_l\rho_r). 
\label{aux1*}
\end{eqnarray} 
Using properties of the projection operator, the first term on the
right-hand side of Eq.~(\ref{aux1*}) can be simplified using the fact that
\begin{eqnarray}
\left\langle F^{\dagger}(t)\right\rangle=
\left\langle e^{t{\mathcal Q}{\mathcal L}}F\right\rangle=
{\mathcal P}e^{t{\mathcal Q}{\mathcal L}}F=
\mathcal P F=
\left\langle F\right\rangle.
\label{aux1**}
\end{eqnarray}
For a homogeneous system, one expects $\langle F \rangle =0$ and hence
$\langle F^{\dagger}(t) \rangle=0$.  For an asymmetric system, on the
other hand, $\langle F \rangle \neq 0$ and the first term on the
right-hand side of Eq.~(\ref{aux1*}) generally does not vanish. 

The second term in the right-hand side of Eq.~(\ref{aux1*}),
involving $\frac{\partial}{\partial X}\rho_l\rho_r$,
can be worked out taking into account that in 
the distributions $\rho_\alpha=Z_\alpha^{-1}e^{-\beta_\alpha H_0^\alpha}$ 
not only the Hamiltonians but also
the partition functions $Z_\alpha$ depend parametrically on $X$,  
\begin{eqnarray}
\frac{\partial}{\partial X}(\rho_l\rho_r)=
\rho_l\rho_r\left(\beta_lF_l+\beta_rF_r-
\frac{1}{Z_l}\frac{\partial Z_l}{\partial X}-
\frac{1}{Z_r}\frac{\partial Z_r}{\partial X}\right).
\label{aux1***}
\end{eqnarray}
It is straightforward to show that
\begin{eqnarray}
\frac{1}{Z_\alpha}\frac{\partial Z_\alpha}{\partial X}=
\beta_\alpha \langle F_\alpha\rangle,
\end{eqnarray}
where $F_l$ and $F_r$ are the forces on the left and right surfaces of the
fixed piston such that $F_l+F_r=F$. 
Therefore, Eq.~(\ref{aux1*}) takes the form
\begin{eqnarray}
\left\langle\frac{\partial}{\partial X}F^\dagger(t)\right\rangle&=&
\frac{d\langle F \rangle}{dX} -\beta_l\langle\!\langle F_lF^\dagger(t)\rangle\!\rangle
-\beta_r\langle\!\langle F_rF^\dagger(t)\rangle\!\rangle,
\end{eqnarray}
where the double brackets denote the cumulants,
$\langle\!\langle AB\rangle\!\rangle=\langle AB\rangle-
\langle A\rangle\langle B\rangle$.
Substituting this result into Eq.~(\ref{f3}), we get
the following exact equation of motion of the piston
\begin{eqnarray}
\frac{dP_*(t)}{dt} &=&\lambda F^\dagger(t) +
{\lambda}^2\int_0^t d\tau\,\,
e^{{\mathcal L}(t-\tau)}\left\{\frac{\partial}{\partial P_*}
\langle FF^\dagger(\tau)\rangle + \frac{P_*}{m} \frac{d\langle F
\rangle}{dX} \right. \nonumber \\
&& \left. -\frac{P_*\beta_l}{m}\langle\!\langle
F_lF^\dagger(\tau)\rangle\!\rangle -
\frac{P_*\beta_r}{m}\langle\!\langle F_rF^\dagger(\tau)\rangle\!\rangle
\right\}.
\label{EM1}
\end{eqnarray}
For some systems, such as the extended Rayleigh model discussed in
Section~4, the parametric dependence of $\langle F \rangle$ on the
position $X$ of the piston is weak, and the derivative of $\langle F
\rangle$ scales as $1/L$ in the limit of large $L$.  In fact for the
Rayleigh model, the bath is comprised of ideal gas molecules, and 
the parametric dependence of
$\langle F \rangle$ on $X$ arises through the dependence of $\langle F
\rangle$ on the concentrations of bath molecules in the respective
compartments of the cylinder, $n_\alpha$.   Since
\begin{eqnarray*}
n_l = \frac{N_l}{(L/2+X)S} \qquad n_r=\frac{N_r}{(L/2-X)S},
\end{eqnarray*}
where $N_l$ and $N_r$ are the total number of bath molecules in the left
and right compartments, respectively, $dn_{\alpha}/dX \sim n_\alpha /
L$ for large $L$, and hence $d\langle F \rangle/dX \sim 1/L$.  Thus,
for large cylinders, this term may be dropped in the equation of
motion of the piston and we obtain
\begin{eqnarray}
\frac{dP_*(t)}{dt} &=&\lambda F^\dagger(t) +
{\lambda}^2\int_0^t d\tau\,\,
e^{{\mathcal L}(t-\tau)}\left\{\frac{\partial}{\partial P_*}
\langle FF^\dagger(\tau)\rangle -\frac{P_*\beta_l}{m}\langle\!\langle
F_lF^\dagger(\tau)\rangle\!\rangle \right. \nonumber \\
&& - \left. \frac{P_*\beta_r}{m}\langle\!\langle F_rF^\dagger(\tau)\rangle\!\rangle
\right\},
\label{EM2}
\end{eqnarray}
which is exact for such systems in the thermodynamic limit.  By
expanding Eq.~(\ref{EM2}) in powers of the square root of the mass
ration $\lambda$, one
can derive the Langevin equation to any order in $\lambda$.

\section{The non-linear Langevin equation}
The force $F^{\dagger}(t)=e^{{\mathcal Q}{\mathcal L}t}F$
in the Eq.(\ref{EM2}) can be expanded in powers of $\lambda$
using  the fact that ${\mathcal P{\mathcal L}}_0=0$ 
and the operator identity (\ref{operators}), to yield
\begin{eqnarray}
F^{\dagger}(t) &=& e^{({\mathcal L}_0+\lambda{\mathcal Q}{\mathcal
L}_1)t}F \nonumber\\
&=&F_0(t)+
\lambda\int_0^t d\tau\, e^{{\mathcal L}_0(t-\tau)}{\mathcal
Q}{\mathcal L}_1 F_0(\tau)+ O(\lambda^2) ,
\end{eqnarray}
where $F_0(t)=e^{{\mathcal L}_0t}F$ is the force exerted by the bath
molecules on the fixed piston. 
Since ${\mathcal L}_1F_0(t)=\frac{P_*}{m}\frac{\partial}{\partial X} F_0(t)$, 
the above equation
takes the form
\begin{eqnarray}
F^{\dagger}(t)=F_0(t)+\frac{\lambda P_*}{m}
\int_0^t d\tau\, \delta G(t,\tau)+ O(\lambda^2),
\end{eqnarray}
where
\begin{eqnarray}
G(t,\tau)=e^{{\mathcal L}_0(t-\tau)}\frac{\partial}{\partial X}F_0(\tau)
\end{eqnarray}
and $\delta G=G-\langle G\rangle$. 
Using this result, one can extract the leading order approximation in $\lambda$
of the correlation functions $\langle FF^\dagger(t)\rangle$ and
$\langle\!\langle F_\alpha F^\dagger (t)\rangle\!\rangle$
appearing in the equation of motion (\ref{EM2}).
Since the evolution of the dynamical functions $F_0(t)$ and $G(t)$ 
is governed by the Liouville operator ${\mathcal L}_0$,
the left and right components of $F_0(t)$ and $G(t)$ are independent,
which implies that the cumulants 
$\langle\!\langle F_l F_0^r\rangle\!\rangle$,
$\langle\!\langle F_r F_0^l\rangle\!\rangle$,  
$\langle\!\langle F_l G_r\rangle\!\rangle$, 
and $\langle\!\langle F_r G_l\rangle\!\rangle$ are all zero.
Taking this into 
account, one obtains
\begin{equation}
\langle\!\langle F_\alpha F^\dagger(t)\rangle\!\rangle=
\langle\!\langle F_\alpha F_0^\alpha(t)\rangle\!\rangle+
\frac{\lambda P_*}{m}\int_0^t dt'
\langle\!\langle F_\alpha G_\alpha(t,t')\rangle\!\rangle,
\nonumber
\end{equation}
and
\begin{equation}
\langle FF^\dagger(t)\rangle=\langle FF_0(t)\rangle+
\frac{\lambda P_*}{m}\int_0^t dt'\Bigl\{
\langle\!\langle F_l G_l(t,t')\rangle\!\rangle+
\langle\!\langle F_r G_r(t,t')\rangle\!\rangle
\Bigr\}. \nonumber
\end{equation}
Substitution of these expressions into the equation of motion (\ref{EM2})
leads to the non-Markovian (generalized) non-linear Langevin equation,
\begin{eqnarray}
\frac{dP_*(t)}{dt} &=& \lambda F^\dagger(t)-
\lambda^2\int_0^t d\tau P_*(t-\tau) M_1(\tau)- 
\lambda^3\int_0^t d\tau P_*^2(t-\tau) M_2(\tau)
\nonumber \\
&&+\lambda^3\int_0^t d\tau M_3(\tau) +O(\lambda^4), 
\label{NMLE}
\end{eqnarray}
where the memory functions are given by
\begin{eqnarray}
M_1(\tau) &=& \frac{1}{m}\Bigl\{
\beta_l\langle\!\langle F_l F_0^l(\tau)\rangle\!\rangle+
\beta_r\langle\!\langle F_r F_0^r(\tau)\rangle\!\rangle \Bigl\},\\
M_2(\tau) &=&\frac{1}{m^2} \int_0^\tau d\tau'\Bigl\{
\beta_l\langle\!\langle F_l G_l(\tau,\tau')\rangle\!\rangle+
\beta_r\langle\!\langle F_r G_r(\tau,\tau')\rangle\!\rangle\Bigr\},\\
M_3(\tau) &=&\frac{1}{m} \int_0^\tau d\tau'\Bigl\{
\langle\!\langle F_l G_l(\tau,\tau')\rangle\!\rangle+
\langle\!\langle F_r G_r(\tau,\tau')\rangle\!\rangle\Bigr\}.
\end{eqnarray}
Note that for a totally symmetric system in which all parameters
characterizing the bath to the left and to the right of the piston 
are the same, 
$\langle\!\langle F_l G_l(\tau,\tau')\rangle\!\rangle=
-\langle\!\langle F_r G_r(\tau,\tau')\rangle\!\rangle$. In this case
the functions $M_2(t),\, M_3(t)$ vanish, and 
the Langevin equation~(\ref{NMLE}) is linear. Furthermore, for a
symmetric bath it can be shown that the first 
non-linear correction term is of order $\lambda^4$ and proportional 
to $P^3$~\cite{PS}.  

Assuming that the memory functions $M_i(t)$ decay with a characteristic 
time $\tau_c$
which is short on the time-scale for relaxation of the momentum
of the piston, the generalized Langevin equation (\ref{NMLE}) can be written 
in 
form that is local in time. In fact, on a time scale  $\tau<\tau_c$ the 
momentum
changes primarily due to the random force while the effect of 
the damping force is of higher order in $\lambda$, 
\begin{eqnarray}
P_*(t-\tau)=P_*(t)-\int_{t-\tau}^t d\tau'\dot{P_*}(\tau')=P_*(t)-
\lambda\int_{t-\tau}^t d\tau'F^{\dagger}(\tau')+{\mathcal O}(\lambda^2).
\end{eqnarray}
Substitution of this expression and a similar one for $P^2(t-\tau)$
into Eq.(\ref{NMLE}) leads to an expression that is local in time and of 
the form,
\begin{eqnarray}
\frac{dP_*(t)}{dt}=\lambda \tilde{F}^{\dagger}(t)-\lambda^2
\zeta_1(t) P_*(t)-
\lambda^3\zeta_2(t) P_*^2(t)
+\lambda^3\zeta_3(t) +{\mathcal O} (\lambda^4),
\label{MLE}
\end{eqnarray}
where $\zeta_i(t)=\int_0^t d\tau\, M_i(\tau)$ and  
$\tilde{F}^{\dagger}(t)$ is the random force 
modified with a correction of second order in $\lambda$
\begin{eqnarray}
\tilde{F}^{\dagger}(t)=
F^{\dagger}(t)-\frac{\lambda^2}{m}\int_0^t d\tau M_1(\tau)
\int_{t-\tau}^t d\tau' F^{\dagger}(\tau').
\end{eqnarray}
Since this correction is small and 
does not change statistical properties of the force, it will be neglected
below.
In what follows the dynamics of the piston on a time-scale longer
than the bath correlation time $\tau_c$ are considered. When $t \gg
\tau_c$, the kinetic coefficients $\zeta_i(t)$ assume their limiting values
$\zeta_i=\int_0^\infty d\tau\, M_i(\tau)$.
Re-expressing this equation in terms of the unscaled momentum of 
the piston $P=P_*/\lambda$ yields
\begin{eqnarray}
\frac{dP(t)}{dt}={F}^{\dagger}(t)-\gamma_1 P(t)-\gamma_2 P^2(t)
+\gamma_3,
\label{LE}
\end{eqnarray}
where the kinetic coefficients $\gamma_i$ are given by 
\begin{eqnarray}
\gamma_1&=&\lambda^2\zeta_1=\frac{1}{M}
\int_0^\infty dt\Bigl\{
\beta_l\langle\!\langle F_l F_0^l(\tau)\rangle\!\rangle+
\beta_r\langle\!\langle F_r F_0^r(\tau)\rangle\!\rangle\Bigr\},\label{gamma1}\\
\gamma_2&=&\lambda^4\zeta_2=\frac{1}{M^2}
\int_0^\infty dt\int_0^t d\tau \Bigl\{
\beta_l\langle\!\langle F_l G_l(t,\tau)\rangle\!\rangle+
\beta_r\langle\!\langle F_r G_r(t,\tau)\rangle\!\rangle\Bigr\},\label{gamma2}\\
\gamma_3&=&\lambda^2\zeta_3=\frac{1}{M}
\int_0^\infty dt\int_0^t d\tau \Bigl\{
\langle\!\langle F_l G_l(t,\tau)\rangle\!\rangle+
\langle\!\langle F_r G_r(t,\tau)\rangle\!\rangle\Bigr\}.
\label{gamma3}
\end{eqnarray}
It is important to note that Eq.~(\ref{LE}) with the kinetic
coefficients given by Eqs. (\ref{gamma1}) - (\ref{gamma3}) is a general
result that is valid for arbitrary interaction potentials for the bath
and that is also independent of the specific form of the interactions
between the piston and the bath molecules.

Note that the correlations 
$\langle\!\langle F_l G_l(t,\tau)\rangle\!\rangle $ and
$\langle\!\langle F_r G_r(t,\tau)\rangle\!\rangle$ are of different sign,
which suggests that for the asymmetric bath either 
$\gamma_2$ or $\gamma_3$  can vanish for certain combinations of the
parameters of the bath. If $\gamma_2=0$, the Langevin equation is
linear and can be easily integrated.
We postpone further discussion of these issues until the next section where
the general results (\ref{LE})-(\ref{gamma3}) 
are applied for the specific model of ideal gas molecules
interacting with the piston via a truncated parabolic potential.

\section{The extended Rayleigh model}
The kinetic constants $\gamma_i$ in Eqs.~(\ref{gamma1})-(\ref{gamma3})
are expressed in terms of integrals of 
correlation functions of dynamical variables whose time evolution is 
described by the Liouvillian operator ${\mathcal L}_0$ of the bath 
in the presence of the fixed piston. These correlations can be calculated
explicitly~\cite{PS} in
the case when the bath is comprised of ideal gas molecules interacting 
with the piston via a parabolic repulsive 
potential
\begin{equation}
U(x-X)= 
\left\{   
\begin{array}{ccl}
\frac{1}{2}k_f\left(x-X_l \right)^{2}&,& x>X_l \\
\frac{1}{2}k_f\left(x-X_r \right)^{2}&,& x<X_r \\
0 &,& {\mbox otherwise},
\end{array}
\right.
\end{equation}
Here $x$ is the coordinate of a bath molecule,
$X_l=X-a$, $X_r=X+a$ define the boundaries of interaction between the
molecule and the piston, with the parameter $a$ serving as a measure 
of the range of the potential. In the model system, the width of 
the piston is neglected.  In addition, it is assumed that
the temperatures of the ideal gas in both compartments are sufficiently 
low (or the potential strength constant
$k_f$ is sufficiently large) that the average penetration 
$(k_f\beta_\alpha)^{-1/2}$ of a bath
particle into the interaction regions $(X_\alpha,X)$ is much less 
than $a$~\cite{approx}.   
 
In the limit of large cylinder length $L$, recollisions of the piston
and gas due to the finite size of the bath can be ignored.  This
assumption corresponds to analyzing the motion of the piston on
intermediate time scales $t \ll \tau_L$, where $\tau_L$ is the time it
takes on average for a bath particle to travel half the length of the
cylinder.  For $t \ll \tau_L$, the force acting on the sides of the
piston can be written in the form~\cite{PS}
\begin{eqnarray*}
F_l(t)&=& -k_f \int_{0}^{\infty} dv \int_{-v\tau_l}^{0} dq
\, N(X_l+q,v;t-\tau_l)\, \frac{v}{\omega_l}\,
\sin{\frac{\omega_l q}{v}} ,\\
F_r(t)&=&-k_f \int_{-\infty}^{0} dv \int_{0}^{-v\tau_r} dq
\, N(X_r+q,v;t-\tau_r)\, \frac{v}{\omega_r}\,
\sin{\frac{\omega_r q}{v}} ,
\end{eqnarray*}
where $\omega_\alpha = \sqrt{k_f/m_\alpha}$ is the characteristic
frequency of the parabolic system, $\tau_\alpha =
\pi/\omega_\alpha$, and the function $N(x,v;t)$ is the microscopic
linear density of particles defined by
\begin{eqnarray*}
N(x,v;t)=\sum_{i} \delta (x-x_i(t)) \delta(v-v_i(t)).
\end{eqnarray*}
The average forces are:
\begin{equation}
\langle F_l \rangle = \frac{n_lS}{\beta_l}  \qquad \langle F_r \rangle
= - \frac{n_rS}{\beta_r}.
\label{forces}
\end{equation}
Note that as mentioned in Section~2, the average force 
$\langle F\rangle =\langle F_l\rangle +\langle F_r\rangle$
does not vanish
in general in the asymmetric system though the derivative 
$d\langle F \rangle/dX \sim
1/L$ can be neglected  for a long cylinder. The equation of motion of the 
piston is
given by Eq.~(\ref{EM2}) in this limit.

For this model, it was found that~\cite{PS}
\begin{eqnarray}
\langle\!\langle F_\alpha F_0^\alpha(t)\rangle\!\rangle
&=&\sqrt{{k_f}/{\beta_\alpha}}\,\,|\langle F_\alpha\rangle|\,\,
\xi_1^\alpha(t),\label{cum1}\\
\langle\!\langle F_\alpha G_\alpha(t_1,t_2)\rangle\!\rangle&=&
-k_f\,\langle F_\alpha\rangle\,\xi_2^\alpha(t_1,t_2)\label{cum2},
\end{eqnarray} 
where the functions $\xi_i^\alpha(t)$ are
\begin{eqnarray}
\xi_1^\alpha(t) &=& \frac{1}{\sqrt{2\pi}}\Bigl\{
\sin\omega_\alpha t+(\pi-\omega_\alpha t)\cos\omega_\alpha t\Bigr\}\,
\theta(\tau_\alpha-t),\\
\xi_2^\alpha(t_1,t_2) &=& \frac{1}{2}
\cos\omega_\alpha t_2(1+\cos\omega_\alpha t_1)
\theta(\tau_\alpha-t_1)\,\theta(\tau_\alpha-t_2),
\end{eqnarray}
and $\theta(t)$ is the step function.

The above results and Eqs. (\ref{gamma1})-(\ref{gamma3})
lead to the following 
expressions for the kinetic coefficients,
\begin{eqnarray}
\gamma_1&=&\sqrt{\frac{8}{\pi}}\frac{1}{M}\Bigl\{
\sqrt{m_l\beta_l}\langle F_l\rangle-
\sqrt{m_r\beta_r}\langle F_r\rangle\Bigr\},\label{gamma21}\\
\gamma_2&=&-\frac{1}{M^2}\Bigl\{
m_l\beta_l\langle F_l\rangle+
m_r\beta_r\langle F_r\rangle
\Bigr\},\label{gamma22}\\
\gamma_3&=&-\frac{1}{M}\Bigl\{
m_l\langle F_l\rangle+
m_r\langle F_r\rangle
\Bigr\}.
\label{gamma23}
\end{eqnarray}
Note that these expressions are independent of the force constant 
$k_f$ of the quadratic potential. 

Since $\langle F_\alpha\rangle =\pm n_\alpha S/\beta_\alpha$ 
one can see from (\ref{gamma22})
that  
$\gamma_2=0$ when 
the mass densities in both compartments 
are the same, $m_ln_l=m_rn_r$. 
In this case the Langevin equation (\ref{LE})
becomes linear and has a solution
\begin{eqnarray}
P(t)=Pe^{-\gamma_1t}+\int_0^t dt e^{\gamma_1(t-\tau)}F^{\dagger}(\tau) 
+\frac{\gamma_3}{\gamma_1}\left(1-e^{-\gamma_1t}\right)
\label{linear_solution}
\end{eqnarray}
which describes relaxation of the momentum to the stationary value
$\langle P\rangle =\langle F\rangle/\gamma_1+\gamma_3/\gamma_1$. 
If $\langle F\rangle=0$ it follows from (\ref{gamma21}) and 
(\ref{gamma23}) that the stationary momentum equals
\begin{eqnarray}
\langle P\rangle=\sqrt{\frac{\pi}{8}}\frac{m_r-m_l}{\sqrt{m_l\beta_l}+
\sqrt{m_r\beta_r}},
\label{station_lin}
\end{eqnarray}
which means that the piston 
moves in the direction of the compartment with the heavier bath particles.
Conditions of equal mass densities, $m_ln_l=m_rn_r$, and 
of equal pressure, $n_l/\beta_l=n_r/\beta_r$, 
are satisfied simultaneously only when $m_l\beta_l=m_r\beta_r$. 
Therefore, Eq.(\ref{station_lin}) describes motion in the direction
of the compartment of higher temperature.

In the general case where $\gamma_2\ne 0$, 
the non-linear Langevin equation~(\ref{LE}) has a form of 
the non-linear Riccati equation and cannot be explicitly 
integrated.  Using the transformation $u=\exp(\gamma_2 P)$  it can be 
converted into a second-order linear equation. In the following, however, 
this line of reasoning will not be pursued and a simple 
perturbation analysis of the non-linear Langevin equation will be conducted.

\section{Stationary and transient solutions} 
Let us first analyze the stationary solution of 
Eq.~(\ref{LE}).  Since 
$\langle F^\dagger(t)\rangle=\langle F\rangle$,
as demonstrated in Eq.~(\ref{aux1**}), the Langevin equation~(\ref{LE}) 
implies that
the stationary value of the momentum of the piston
is given by
\begin{eqnarray}
\langle P\rangle=\frac{1}{\gamma_1}\,\langle F\rangle-
\frac{\gamma_2}{\gamma_1}\,\langle P^2\rangle +
\frac{\gamma_3}{\gamma_1}.
\label{station1}
\end{eqnarray}
To calculate $\langle P\rangle$ perturbatively to first order in
$\lambda$, one substitutes in Eq.~(\ref{station1}) the lowest in $\lambda$
approximations for $\langle P^2\rangle$, which can be derived solving
the linear Langevin equation 
\begin{equation}
\dot P(t)=F_0(t)-\gamma_1 P(t).
\label{OLE}
\end{equation}
Note that here the ``random'' force $F_0$  generally is not 
zero-centered. In the long time limit, 
the auto-correlation functions for the force on the piston in the left 
and right 
compartments are effectively
\begin{eqnarray}
\langle F_\alpha F_0^\alpha(t)\rangle=
\langle F_\alpha\rangle^2+
\langle\!\langle F_\alpha F_\alpha^0(t)\rangle\!\rangle \to
\langle F_\alpha\rangle^2+2\Gamma_\alpha\delta(t),
\end{eqnarray}
where
$\Gamma_\alpha=\int_0^\infty dt\,
\langle\!\langle F_\alpha F_\alpha^0(t)\rangle\!\rangle $. 
Thus, the  correlation function of the total force 
$F_0=F_0^l+F_0^r$ is
\begin{eqnarray}
\langle FF_0(t)\rangle=
\langle F\rangle^2+2\Gamma\delta(t),
\label{corr}
\end{eqnarray}
where $\Gamma=\Gamma_l+\Gamma_r$.

Using~(\ref{corr}), one can calculate $\langle P^2(t)\rangle$ 
from the solution of the linear Langevin equation~(\ref{OLE}), namely,
\begin{eqnarray}
P_0(t)=Pe^{-\gamma_1t}+\int_0^td\tau\,e^{-\gamma_1(t-\tau)}F_0(\tau).
\label{P0}
\end{eqnarray}
For $t\gg 1/\gamma_1$, the result assumes the form  
\begin{equation}
\langle P_0^2\rangle=\frac{\Gamma}{\gamma_1}+
\frac{1}{\gamma_1^2}\langle F\rangle^2 ,
\label{P2}
\end{equation}
where $\Gamma$ is given by
\begin{eqnarray}
\Gamma=\int_0^\infty dt\,
\langle\!\langle F_l F_l^0(t)\rangle\!\rangle+
\int_0^\infty dt\,
\langle\!\langle F_r F_r^0(t)\rangle\!\rangle.
\end{eqnarray}

It has been assumed throughout our analysis that the contribution of the 
systematic force $\langle F\rangle$ to
the  momentum of the piston is small. This implies that
the second term on the right hand side of Eq.~(\ref{P2}) is small
compared to the first one. Since $\Gamma\sim\lambda^0$ and 
$\gamma_1\sim\lambda^2$ this restriction requires that 
$\langle F\rangle\sim\lambda^{1+\epsilon}$, for some $\epsilon>0$.

For the truncated quadratic potential, $\Gamma$ can be evaluated
using Eq.~(\ref{cum1}) to obtain 
\begin{eqnarray}
\Gamma=\sqrt{\frac{8}{\pi}}\left\{
\sqrt{\frac{m_l}{\beta_l}}\langle F_l\rangle-
\sqrt{\frac{m_r}{\beta_r}}\langle F_r\rangle\right\}.
\end{eqnarray}
It then follows that for the case $\langle F \rangle = 0$, one obtains 
\begin{eqnarray}
\langle P_0^2\rangle=M\frac
{\sqrt{m_l/\beta_l}+\sqrt{m_r/\beta_r}}
{\sqrt{m_l\beta_l}+\sqrt{m_r\beta_r}}.
\label{P_square}
\end{eqnarray}
For a random force obeying Gaussian statistics, 
distribution of the momentum of the piston is of Maxwellian form
with the effective temperature $k_B T_p=\langle P_0^2\rangle/M$. 
In particular, if $m_l=m_r$, then it is found that
$T_p=\sqrt{T_lT_r}$ from Eq.~(\ref{P_square}),
in agreement with the results of Gruber and Piasecki~\cite{GP}.

Substitution of expression~(\ref{P2}) 
for $\langle P_0^2\rangle$ into Eq.~(\ref{station1}) yields
for the stationary momentum,
\begin{eqnarray}
\langle P\rangle=\frac{1}{\gamma_1}\,\langle F\rangle-
\frac{\gamma_2}{\gamma_1^3}\,\langle F\rangle^2+
\frac{\gamma_3}{\gamma_1}
-\frac{\gamma_2\Gamma}{\gamma_1^2}.
\label{station2}
\end{eqnarray}
For $\langle F\rangle=0$ and 
with  explicit expressions (\ref{gamma21})-(\ref{gamma23})  for $\gamma_i$,
Eq. (\ref{station2}) takes the form
\begin{eqnarray}
\langle P\rangle=\sqrt{\frac{\pi}{8}}
\frac{ \sqrt{m_lm_r}}{ \sqrt{m_l\beta_l}+\sqrt{m_r\beta_r} }
\left(
\sqrt{\frac{\beta_l}{\beta_r}}-\sqrt{\frac{\beta_r}{\beta_l}}\right).
\label{station3}
\end{eqnarray}
When the masses of the bath particles are the same in both compartments,
$m_l=m_r\equiv m$, this expression reduces to the result of  
Gruber and Piasecki~\cite{GP}
\begin{eqnarray}
\langle P\rangle=\sqrt{\frac{\pi}{8}}\sqrt{m}\left(\sqrt{k_BT_r}-\sqrt{k_BT_l}
\right).
\end{eqnarray}
Note that in this case, $\gamma_3$ vanishes  as can be seen
from Eq.~(\ref{gamma23}). If $m_l\beta_l=m_r\beta_r$  
(and therefore $\gamma_2=0$),
Eq.~(\ref{station3}) coincides with the result (\ref{station_lin})
obtained in the previous section.

In addition to the stationary solutions of the non-linear Langevin
equation, perturbative solutions of the time evolution of the momentum
of the piston can be examined
by expanding the momentum in the form $P(t)=P_0(t)+P_1(t)+\cdots$
where $P_0(t)$ is the solution (\ref{P0}) of the linear Langevin equation
~(\ref{OLE}) and $P_1(t)$ is the next-order correction in $\lambda$ which 
satisfies
the equation
\begin{eqnarray}
\frac{dP_1(t)}{dt}=F_1(t)-\gamma_1 P_1(t)-\gamma_2 P_0^2(t)
+\gamma_3,
\label{LLE}
\end{eqnarray}
and the condition $P_1(0)=0$.
In this equation, the force $F_1(t)=F^{\dagger}(t)-F_0(t)$ is zero centered,
since $\langle F^{\dagger}(t)\rangle=\langle F_0(t)\rangle=
\langle F(0)\rangle$, as established in Eq.~(\ref{aux1**}).
Solving Eq. (\ref{LLE}) and taking the 
average gives
\begin{eqnarray}
\langle P_1(t)\rangle=
&&\frac{\gamma_3}{\gamma_1}\left(1-e^{-\gamma_1t}\right)
-\frac{\gamma_2}{\gamma_1}P^2e^{-\gamma_1t}\left(1-e^{-\gamma_1t}\right)
\nonumber\\
&&-\frac{2\gamma_2}{\gamma_1^2}\langle F\rangle Pe^{-\gamma_1t}
\Bigl\{\gamma_1t-\left(1-e^{-\gamma_1t}\right)\Bigr\} \nonumber\\
&&-\frac{2\gamma_2}{\gamma_1^2}\langle F\rangle^2 e^{-\gamma_1 t}
\left(\sinh\gamma_1t-\gamma_1t\right) \nonumber\\
&&-\frac{2\gamma_2\Gamma}{\gamma_1^2}e^{-\gamma_1t}\left(\cosh\gamma_1t-1
\right),
\end{eqnarray}
where $P=P(0)$.
Note that the average momentum
$\langle P\rangle =
\langle P_0(t)\rangle+\langle P_1(t)\rangle$
takes the stationary value (\ref{station2}) in the long-time limit $t\gg 
\gamma_1^{-1}$.  It should be emphasized that 
the transient solution obtained here is valid only on a time scale
that is much longer than the characteristic time $\tau_c$ for the
relaxation of the bath but shorter than $\tau_L$, the time scale for
particles to move half the length of the cylinder.
The above analysis may be readily extended to the short-time domain
using as a starting point the Langevin equation (\ref{MLE})
with time-dependent damping coefficients.

\section*{Acknowledgments}
This work was supported by a grant from the
Natural Sciences and Engineering Research Council of Canada.

%\end{multicols}


\begin{thebibliography}{9}
\bibitem{correction} It is shown in Ref.~\cite{PS} that for 
the case of the piston in a bath of ideal gas particles, 
$\lambda$ appears in the expansion of the generalized Langevin
equation via the parameter $\lambda_\star=\sqrt{N}\lambda$,
where $N$ is a number of bath particles in the interaction shell
around the piston (provided $N\gg1$). 
The actual parameter controlling separation of time scales
in this case is therefore $\lambda_\star$.
\bibitem{approx} This assumption justifies the parabolic potential
model and is similar to that used in the linear 
theory of lattice vibrations when
displacements of atoms are assumed to be
small compared to the lattice spacing.
\end{thebibliography}
\end{document}